\documentclass[review]{elsarticle}

\usepackage{lineno,hyperref,epsfig}
\modulolinenumbers[5]

\journal{Journal of \LaTeX\ Templates}

\newcommand{\erel}{E_{\rm rel}}
\newcommand{\ex}{E_{\rm x}}
\newcommand{\sn}{S_{\rm n}}
\begin{document}

\begin{frontmatter}

\title{Large Acceptance Spectrometers
for Invariant Mass Spectroscopy of Exotic Nuclei and Future Developments}

\author{T. Nakamura}
\address{2-12-1 O-Okayama, Meguro, Tokyo 152-8551, Japan}

\author{Y. Kondo}
\address{2-12-1 O-Okayama, Meguro, Tokyo 152-8551, Japan}

\begin{abstract}
Large acceptance spectrometers at in-flight RI separators 
have played significant roles in investigating the structure of exotic nuclei.
Such spectrometers are in particular useful for probing unbound states of exotic nuclei,
using invariant mass spectroscopy with reactions at intermediate and high energies.
We discuss here the key characteristic features of such spectrometers, 
by introducing the recently commissioned SAMURAI facility at the RIBF, RIKEN. 
We also explore the issue of cross talk in the detection
of multiple neutrons, which has become crucial for exploring further unbound states and
nuclei beyond the neutron drip line.
Finally we discuss future perspectives for large acceptance spectrometers
at the new-generation RI-beam facilities.
\end{abstract}


\end{frontmatter}

%\linenumbers

\section{Invariant mass spectroscopy in RI-beam experiments}

Rare isotope beams available at in-flight separators, at
RIKEN (RIBF), MSU, GSI, and GANIL, have
expanded physics opportunities to a wider range of $N-Z$.
Accordingly, more experiments have been 
performed for nuclei near the drip line or even beyond~\cite{TANI13,BAUM12}. 
For such nuclei, most or all of the states are 
unbound (i.e., in the continuum), thereby decaying by emitting particles.
The invariant mass spectroscopy of such unbound states produced
with direct reactions and fragmentation of exotic nuclei at intermediate and high energies 
has thus become a powerful experimental tool in RI-beam physics.
Large-acceptance spectrometers play a major role, as we will show, 
in performing invariant mass spectroscopy experiments.

%Invariant mass spectroscopy for unbound states works in the following way.
Let us take an example of the recent experiment on the 1$n$ knockout reaction
$^{17}$C$+p$ at 70 MeV/u by Satou {\it et al.}~\cite{SATO14} at the RIPS
facility~\cite{KUBO92} at RIKEN, where unbound states of $^{16}$C were studied.
In this case, decay particles $^{15}$C and a neutron 
emitted in the forward kinematical cone were measured.
From the momentum vectors of these two particles, one can reconstruct
the invariant mass $M_{16*}$ of a $^{16}$C state as,
\begin{equation}
M_{16*}= \sqrt{ (E_{15}+E_n)^2 - |\vec{P_{15}} - \vec{P_n}|^2},
\end{equation}
where $(E_{15},\vec{P_{15}})$ and $(E_n,\vec{P_n})$ are the four momenta of the
$^{15}$C fragment and the neutron. One can extract the relative energy 
$\erel$ and the excitation energy $\ex$ as,
\begin{eqnarray}
\erel=M_{16*}- \left(M_{15}+M_n \right), \\
\ex = \erel+\sn,
\end{eqnarray}
where $M_{15}, M_n$ are the masses of $^{15}$C and the neutron,
and $\sn$ is the neutron separation energy (for $^{16}$C, $\sn$=4.25 MeV).
If the $^{15}$C is produced
in a bound excited state, 
then the $\gamma$ decay energy ($E_\gamma= 740$~keV for $^{15}$C)
should also be measured and $\ex$ 
is shifted up by $E_\gamma$, since $M_{15}$ is replaced by $M_{15}+E_\gamma$. 
In the experiment, three states were found
at $\erel=$0.46(3), 1.29(2), and 1.89 MeV that correspond to $\ex=$5.45(1), 6.28(2) and 6.11 MeV.
The 6.28(2) MeV state was found to be in coincidence with the 740~keV $\gamma$ ray, and $\ex$ is shifted accordingly.

The advantages of invariant mass spectroscopy in the study of exotic nuclei
are summarized as follows.
\begin{itemize}
\item Good energy resolution: One can reach an energy resolution of about 
a few hundred keV ($1\sigma$) at $\erel=1$ MeV even for a momentum resolution of the order
of 1\% \cite{AUMA13} for the fragment and neutron individually. 
Note that the relative-energy resolution $\Delta \erel$ 
follows approximately $\Delta \erel \propto \sqrt{\erel}$.
\item Kinematic focusing: Since the outgoing particles are boosted by the beam velocity
at intermediate and high energies, they are emitted in a narrow kinematical cone. 
Consequently, one can detect the decay particles with high geometrical efficiency.
\item Thick target: Since one uses intermediate and high energy beams, one
can use a comparatively thick target of the order of 100 mg/cm$^2$ at 50-70 MeV/u
to 1 g/cm$^2$ at 200 MeV/u. 
% revised
Hence, one can obtain high reaction yield, which
is important for RI-beam experiments since beam intensity is generally week.
\end{itemize}
Owing to these advantages invariant mass spectroscopy has become 
one of the most useful methods to study the continuum structure of exotic nuclei.
There is, however, one disadvantage: One 
needs to measure all the outgoing beam-velocity particles, which makes the experiment and the analysis 
more complicated. For instance, if the daughter nucleus is in a 
high-lying excited state, then this
may decay by a cascade of $\gamma$ rays. In this case, an accurate measurement of
the excitation energy requires a high-efficiency $\gamma$-ray calorimeter.

To realize invariant mass spectroscopy, a large acceptance spectrometer is highly desirable.
In the above example~\cite{SATO14}, a simple dipole magnet was used in combination 
with the neutron-detector array based on plastic scintillators (see Fig.~1 of
Ref.~\cite{NAKA06}, the ``RIPS-Dipole setup''), which 
was a pioneering invariant-mass-spectroscopy setup 
at the RIPS facility at RIKEN since 1992. This dipole magnet 
has a relatively large gap (30 cm), so that the outgoing particles including neutrons
have a large acceptance. On the other hand, the momentum resolution is
moderate (1\%) since focusing elements such as quadrupole magnets are not used.
% revised
A momentum resolution of 1\% is already sufficient to obtain
a good $\erel$ resolution, and a simple dipole magnet has an advantage 
of having large acceptance. 
The use of such a magnet 
is also necessary to ``sweep'' the charged particles away from the neutron detectors.

A large momentum acceptance of the magnet is advantageous 
in studying a variety of final states with a single setup. 
Let us consider the incident beam of the drip-line nucleus $^{22}$C 
on a carbon target.
% revised
In this case, one can study its reaction cross section of $^{22}$C 
to study its size, 1$n$ removal to study the unbound $^{21}$C states 
($\rightarrow ^{20}$C$+n$), low-lying excited states $^{22}$C 
with the inelastic scattering, and other unbound states such as $^{16,17,18,19}$B
with proton-removal fragmentation reactions, for example.

\section{SAMURAI Facility at RIBF}

\begin{figure}[htb]
\begin{center}
\begin{minipage}[ht]{11.5 cm}
\epsfig{file=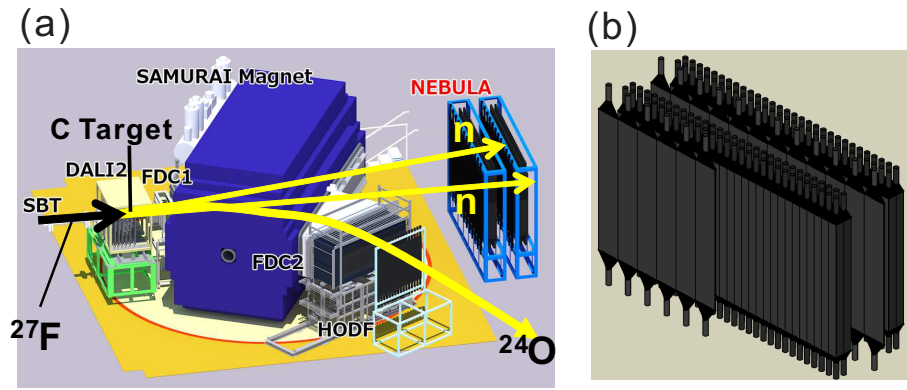,scale=0.6}
\end{minipage}
\begin{minipage}[ht]{10. cm}
\caption{(a) The SAMURAI setup for the invariant mass 
spectroscopy of neutron-rich nuclei,
as used for the study of the unbound states of $^{26}$O~\cite{KOND15}.
(b) The NEBULA neutron detector array. The first veto layer is
partially removed for display purposes.
\label{fig:samurai_nebula}}
\end{minipage}
\end{center}
\end{figure}

At the RIBF, RIKEN, the advanced 
invariant-mass-spectrometer setup, SAMURAI was constructed 
and commissioned in 2012~\cite{KOBA13,SHIM13,SATO13}.
SAMURAI stands for {\bf S}uperconducting {\bf A}nalyser for 
{\bf MU}lti particles from {\bf RA}dio {\bf I}sotope Beams.
The SAMURAI setup for the invariant mass spectroscopy of neutron-rich nuclei
is schematically shown in Fig.~\ref{fig:samurai_nebula}(a). This setup was
used, as shown, for the recent kinematically complete measurement
of the unbound system $^{26}$O
by 1$p$ knockout from $^{27}$F 
with a carbon target at 201 MeV/u~\cite{KOND15}.

\begin{figure}[htb]
%\epsfysize=9.0cm
%\begin{center}
\begin{minipage}[ht]{11.5 cm}
\epsfig{file=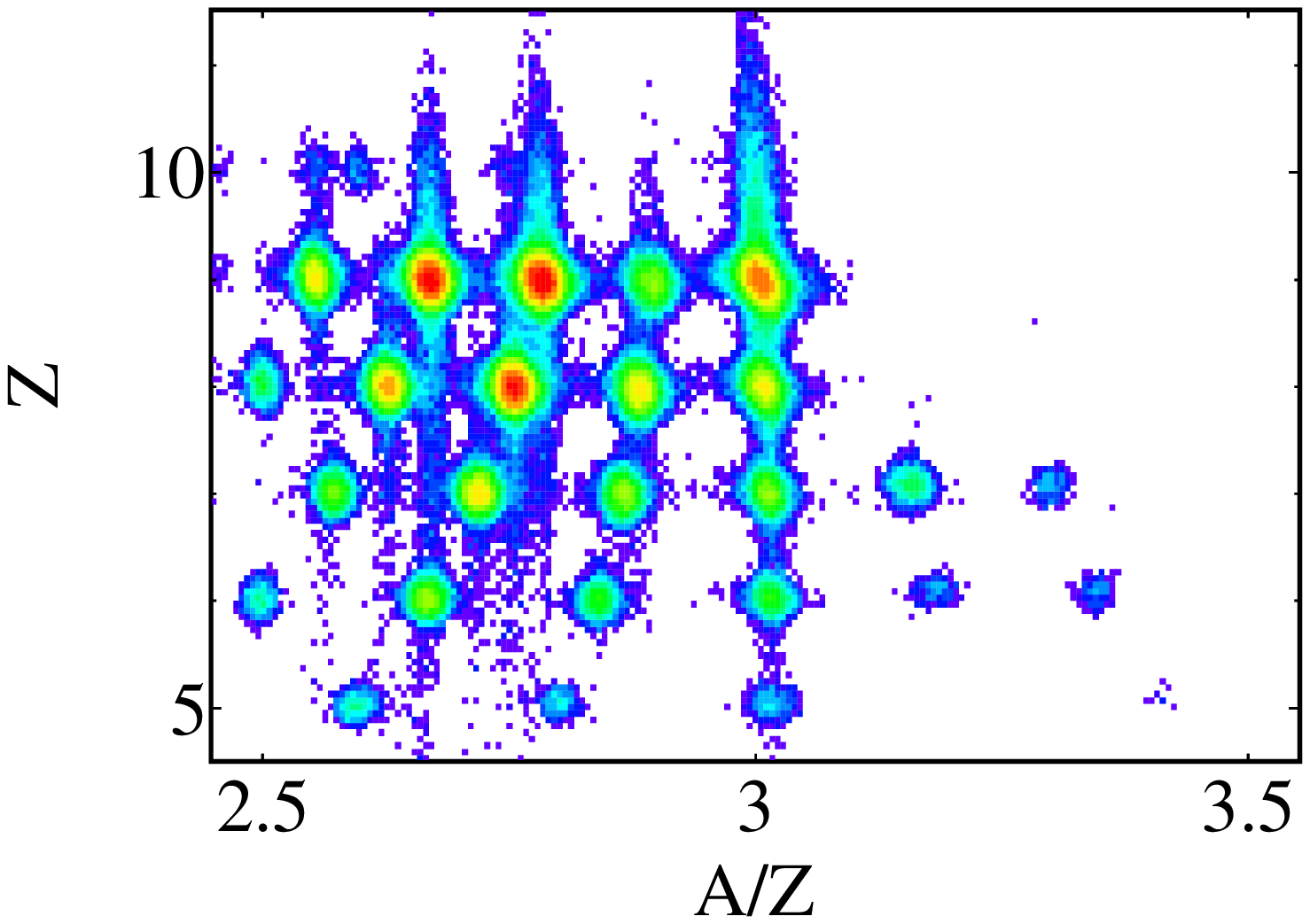,scale=0.38}
\epsfig{file=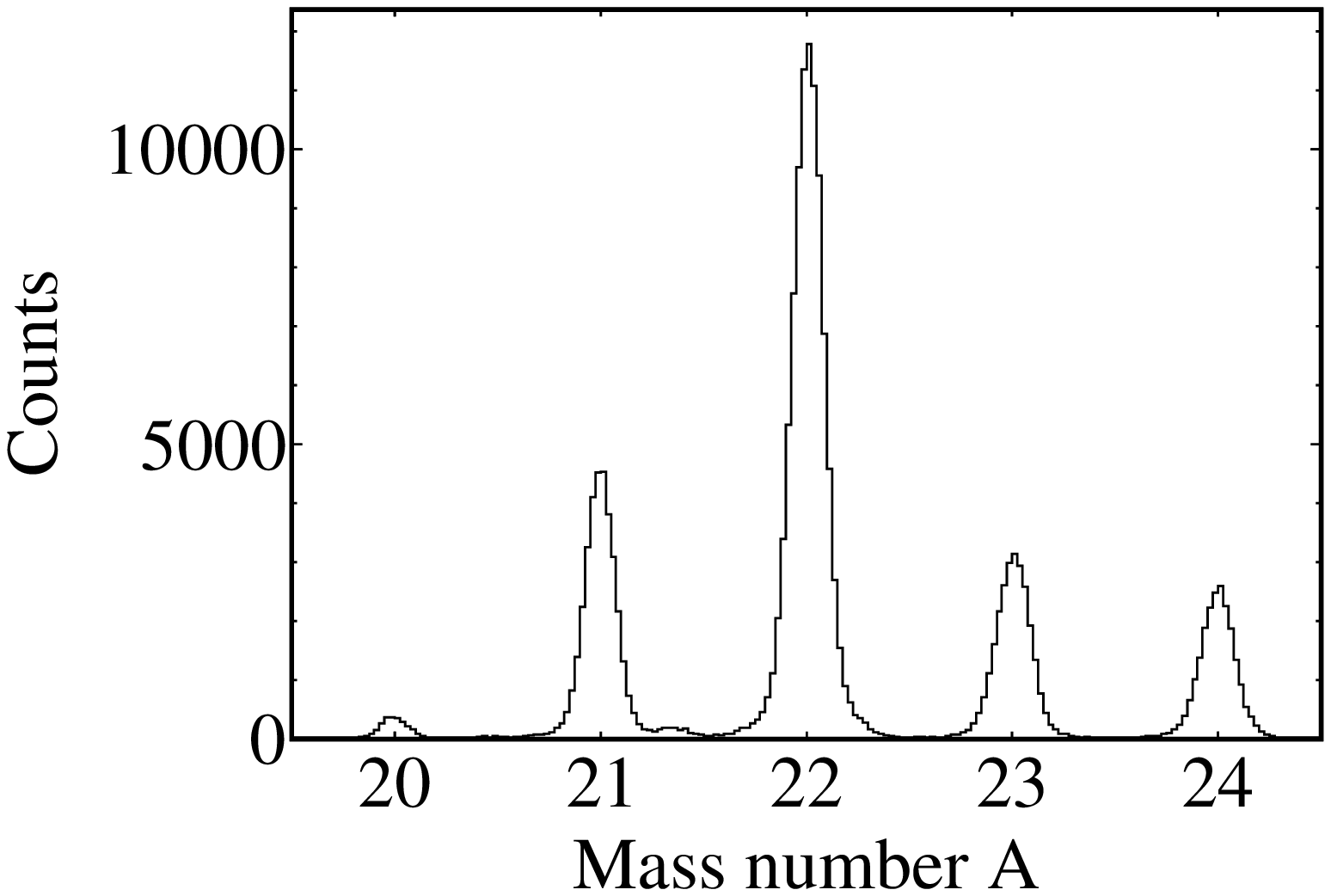,scale=0.38}
\end{minipage}
\begin{center}
\begin{minipage}[ht]{10. cm}
\caption{Left: particle identification spectrum of the charged fragments 
for $^{27}$F$+$C at 201 MeV/u,
obtained from the tracking and the TOF between the target and the hodoscope (HODF).
Right: The mass spectrum of the oxygen isotopes.
\label{fig:pid}}
\end{minipage}
\end{center}
\end{figure}

The principal element 
is the superconducting SAMURAI magnet with a maximum field of 3.1 Tesla
(Field integral 7.1 Tm) with a large effective gap of 80 cm. 
One characteristic feature of the SAMURAI facility is its relatively 
high momentum resolution for the charged fragment, of the
order of 10$^3$ ($1\sigma$).
This was realized by designing the magnet to have a 
large bending angle of about 60~degrees,
as well as the tracking using four multi-wire drift chambers
with high position resolutions~\cite{KOBA13}. 
A simple tracking analysis using a polynomial fit and
the calculated field map, combined with 
a time-of-flight measurement between the target and the hodoscope (HODF),
can already provide $P/\Delta P\sim 700$ ($\sigma$),
the design value of SAMURAI. With detailed tracking and restricted
acceptance, the momentum 
resolution can reach about 1500~\cite{KOBA13}. 
The interest of high-momentum resolution
is that it provides for high mass resolution in the particle-identification.
When one needs sufficient separation in the mass
distribution $\sim5\sigma$ separation may be necessary when a particular isotope
has a much larger yield compared to the neighbors.
Such a high separation (5$\sigma$)
is indeed achieved for charged fragments with $A\sim 100$ 
when the momentum resolution is  $P/\Delta P=700$.
Figure~\ref{fig:pid}~(left) shows the particle identification spectrum obtained
in the $^{26}$O experiment.
The mass spectrum extracted for the oxygen isotopes is shown in Fig.~\ref{fig:pid}~(right), 
where better than $\sim$10$\sigma$ separation is reached in this
mass region.
Recently, an experiment on $^{132}$Sn was performed where masses are clearly 
separated even in this mass region~\cite{YASU15}.

Neutrons emitted in the forward direction go through the gap of the magnet and their positions
and time-of-flight are measured by the neutron detector array NEBULA 
({\bf NE}utron-detection system for {\bf B}reakup of {\bf U}nstable-Nuclei with
{\bf L}arge {\bf A}cceptance), which is shown schematically in Fig.~\ref{fig:samurai_nebula}(b).  
The NEBULA array consists of 120 modules of plastic scintillator,
each of which is 12(W)$\times$12(D)$\times$180(H) cm$^3$. These modules are
 arranged into two walls, each of which is composed of 
two layers of 30 modules. The total thickness
is thus 48~cm~\cite{KOBA13}, and the area amounts to 360$\times$180~cm$^2$. 
In the $^{26}$O experiment, the front faces of these two walls 
were 11.12 m  and 11.96 m downstream of the reaction target. 
Each wall is equipped with a charged-particle veto array of 1~cm thickness.
A wide acceptance is required since neutrons are emitted with much 
larger angles than the charged fragment, as discussed below.

%, which is realized by such large area of NEBULA, combined with
%the large gap of the SAMURAI magnet. The typical 
%acceptance for neutrons is  $\theta_H \pm$8.8$^\circ$, $\theta_V \pm 4.4^\circ$,
%where about 50\% of the events are covered even at $\erel=$5 MeV.

The other important feature of SAMURAI as an advanced large-acceptance facility
is that it offers a variety of experimental modes,
which are owing to the rotatable stage on which the magnet is installed.
The range of rotation is -5$^\circ$ to 95$^\circ$ 
degrees (0$^\circ$ corresponds to the setup where the
entrance and exit faces are 90$^\circ$ to the beam axis).
The setup in Fig.~\ref{fig:samurai_nebula}(a) is at 30$^\circ$.
SAMURAI thus offers a variety of experimental setups, e.g., for 
1) Invariant mass spectroscopy by HI(Heavy-ion fragment) + neutron(s) coincidences 
as in the example of $^{26}$O, 2) Invariant mass spectroscopy by HI+proton coincidence
at the 90$^\circ$ setting, where the hole in the yoke is used as a beam port, 
3) Missing mass spectroscopy by measuring recoil
particles primarily from the target, 4) Polarized deuteron-induced reactions, 
and 5) Heavy-ion collisions to measure $\pi^\pm$ using the 
TPC(Time Projection Chamber) ~\cite{SHAN15}
in the gap of the magnet at the 0$^\circ$ setting. 
As such, SAMURAI is a unique facility capable of supporting a very versatile
nuclear physics program.

\section{Large-acceptance spectrometers vs. High-resolution spectrometers}

As with the SAMURAI facility, 
large-acceptance spectrometers have 
been constructed at many in-flight RI-beam facilities,
and played significant roles in the spectroscopy of unstable nuclei.
At RIKEN, as mentioned, before SAMURAI was commissioned, a smaller RIPS-Dipole setup
had been used. 
At the NSCL at MSU, the Sweeper superconducting magnet
is installed~\cite{BIRD05}, combined with the large acceptance neutron array MoNA
and LISA~\cite{BAUM05}. At GSI, the ALADIN/LAND setup has long been used, and
it is now being upgraded to the R$^3$B setup for the FAIR facility~\cite{R3B}.

The characteristic features of the large-acceptance spectrometers 
are now discussed, in comparison with the high-resolution spectrometers.
Table~\ref{tab:spectro} compares the characteristic features.
The momentum resolution ($P/\Delta P$) is 
of the order of $10^2-10^3$ for the large acceptance spectrometers, 
while that is the order of $10^4$ for the high resolution spectrometers.
The large acceptance spectrometer is intended 
primarily for invariant mass spectroscopy, while the high resolution spectrometer is
for missing mass spectroscopy which requires higher momentum resolution.

  \begin{table}[ht]
    \caption{Comparison of large-acceptance and high-resolution
spectrometers. RIPS-Dipole setup and SAMURAI represent 
the large acceptance spectrometers, while SHARAQ and the S800 Spectrograph
represent the high resolution spectrometers. The angular acceptance for
the large-acceptance spectrometer is for neutrons, while that for the high-resolution
spectrometer is for the charged particle residue (ejectile).
%The momentum acceptance for SHARAQ
%has two values: one is the high resolution mode, and 
%the other, in parenthesis, is the moderate resolution mode.
}
    \label{tab:spectro}       % Give a unique label
    \begin{tabular}{p{2cm}p{2.5cm}p{2.5cm}p{2.5cm}p{2.5cm}}
      \hline\noalign{\smallskip}
      
      & RIPS-Dipole & SAMURAI~\cite{KOBA13} & SHARAQ~\cite{UESA12} & S800~\cite{BAZI03} \\
      \noalign{\smallskip}\hline\noalign{\smallskip}
$P/\Delta P$ & $\sim$ 100 & $\sim$1000 & 15000  & 10000 \\
Angular Acceptance & $\sim$100 mstr  & $\sim$50 mstr & 4.8 mstr & 20 mstr \\
Momentum Acceptance & $\sim$50 \%  & $\sim$50 \%  & 2\% & $\sim$5\% \\
$B\rho_{\rm MAX}$ &  $\sim$4.2 Tm & $\sim$7 Tm & 6.8 Tm & 4 Tm \\
Configuration & D  & D & QQDQD & QQDD  \\
      \noalign{\smallskip}\hline\noalign{\smallskip}
    \end{tabular}
  \end{table}

It is worth noting that high acceptance is needed for the
invariant mass spectroscopy for exotic nuclei for the sake of the
neutron (proton) detection.
For instance, let us consider the invariant mass spectroscopy of $^{A}Z$ 
breaking up into $^{A-1}Z +n$.
In this case, it is easily shown that the emission angle for neutron, $\theta_n$, 
is roughly $A-1$ times the angle $\theta_f$ for the charged fragment,
due to the momentum balance in the center-of-mass frame. 
Hence, the acceptance is more crucial for neutron detection.
The opening angle $\theta$ between the neutron and the fragment 
is then close to $\theta_n$. 
The relative energy is approximately
\begin{equation}
\erel = \frac{1}{2}\mu v_{\rm rel}^2 \sim\frac{E}{A}\theta_n^2,
\end{equation}
where $E$ is the incident beam energy.
This simple consideration demonstrates that when the neutron detectors and the
gap of the magnet allow a measurement of the neutrons up to $\theta_n = 5^\circ (10^\circ)$, then
events of $\erel\simeq 2$ (8) MeV are fully accepted.

%One notable feature is that the recently-commissioned SAMURAI, 
%which may be categorized as the new-generation large acceptance spectrometer, 
%has a resolution of $\sim 10^3$, about one-order better than the conventional large acceptance
%spectrometers.

%\section{Invariant mass setup of SAMURAI}

%The charged fragment ($^{24}$O) 
%is bent by the SAMURAI superconducting magnet ($\sim$3 Tesla) to be tracked
%by two multi-wire drift chambers (FDC1, FDC2)
%at the entrance and exit of the magnet, and its Time-of-Flight(ToF) 
%and $\delta E$ are measured in the plastic-scintillator hodoscope (HODF).
%The details of these SAMURAI standard detectors can be seen in Ref.~\cite{KOBA13}.

%Rare-isotope beam particles ($^{26}$F in this example) 
%produced from an intense heavy ion beam (c.a. 140 pnA $^{48}$Ca at 345 MeV/u), 
%selected and identified at the BigRIPS in-flight separator, are transported
%onto the reaction target (1.8g/cm$^2$-thick carbon) at the SAMURAI facility.
%The tracking by two multi-wire drift chambers (BDC1 and BDC2) determines
%the angle and position of the beam particle onto the target. Unbound states
%($^{26}$O) then decay into an charged fragment and neutrons ($^{24}$O$+n+n$).
%

\section{Neutron detection and cross talk rejection at SAMURAI/NEBULA}

The NEBULA array is, as with the other high-energy 
% revised
neutron detector arrays such as MoNA~\cite{BAUM05} and NeuLAND~\cite{R3B},  
based on plastic scintillator. 
The performance of the NEBULA array was investigated, using the simulation code
GEANT4 with the QGSP\_INCLXX physics model (intranuclear cascade model)
for the neutron interactions in NEBULA. 
The simulation was then compared with the experimental results using the
$^7$Li$(p,n)$$^7$Be reaction at 200 MeV
where the ground and the 1st excited states of $^{7}$Be were populated. 
This reaction can thus deliver nearly mono-energetic
neutrons, and thus has long been used for the evaluation of the characteristics
of neutron detectors.
From the simulation, we found that the intranuclear cascade model used here reproduces
the exprimental results rather well as shown below 
at energies around 200 MeV. 
We note that 
neutron detection below 100 MeV is well understood with MENATE\_R~\cite{ROED08,KOHL12}, 
an updated version of MENATE~\cite{DESQ91}.
%However, we found that
%MENATE\_R is not applicable in this energy range as it is.
%The simulation evaluated the $1n$ detection efficiency at NEBULA as 37\%, 
%while the experiment extracts the efficiency as 34.7$\pm$0.4(stat)$\pm$1.0\%
%with the detection threshold of $E_{\rm th} =6$(MeVee, ee:electron equivalent),
%which is close to the simulated value.

In the invariant mass spectroscopy of neutron-rich nuclei, coincidence detection 
of more than one neutron becomes more imporatnt.
For instance, in the Coulomb breakup of two-neutron halo nuclei, such as $^{11}$Li,
one needs to measure $^9$Li$+n+n$~\cite{NAKA06}. In the study of
$^{26}$O, one needed to measure $^{24}$O+$n$+$n$. 
In the near future, the challenge of detecting four neutrons in coincidence
will need to be confronted for the study of $^{28}$O.
In such cases, one needs to eliminate so-called ``cross talk'', where one neutron
can produce more than one signal that may mimic multi-neutron events.
Such cross-talk events  can be investigated using 
the $^7$Li($p,n$)$^7$Be(g.s.+ 0.43MeV) reaction that emits only a single
neutron. 
As such, all the multiplicity-greater-than-one events in the NEBULA array 
are judeged as cross-talk events. 
%The NEBULA array was installed in two-wall
%configurations: the front face of the first (second) wall is located at xxx cm, and
%that of the second wall is located at yyy cm.
%there is a gap of ?? cm between the read face and the front face of the 1st and 2nd wall.

Here, we consider primarily how to treat the
two-neutron coincidence events to distinguish them from the cross talk.
There are two ways to detecting two neutrons in an array such as NEBULA: 
i) Different-wall events:
one neutron detected in the 1st wall and the other neutron in the 2nd wall, ii)
Same-wall events: both of the neutons are detected 
in the same wall, either in the 1st or 2nd wall.
We discuss these two cases separately.
We note that cross-talk rejection procedures have also been
developed at lower energies~\cite{Wan97,Mar00}.

\subsection{Cross talk in different-wall events}

\begin{figure}[htb]
\begin{center}
\begin{minipage}[ht]{10. cm}
\epsfig{file=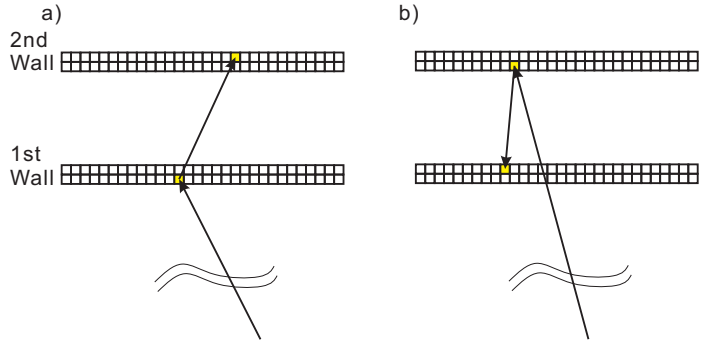,scale=0.5}
\end{minipage}
\begin{minipage}[ht]{10. cm}
\caption{Examples of cross-talk events relevant to the different wall events.
(a) A single neutron is scattered and leaves a signal in a module in the 1st Wall,
then leaves another signal in a module in the 2nd Wall.
(b) A single neutron is registered in a scintillator in the 2nd wall, and the evaporated
neutron is detected in the 1st Wall.
\label{fig:cross_df_nebula}}
\end{minipage}
\end{center}
\end{figure}

\begin{figure}[htb]
%\epsfysize=9.0cm
%\begin{center}
\begin{minipage}[htb]{11.5 cm}
\epsfig{file=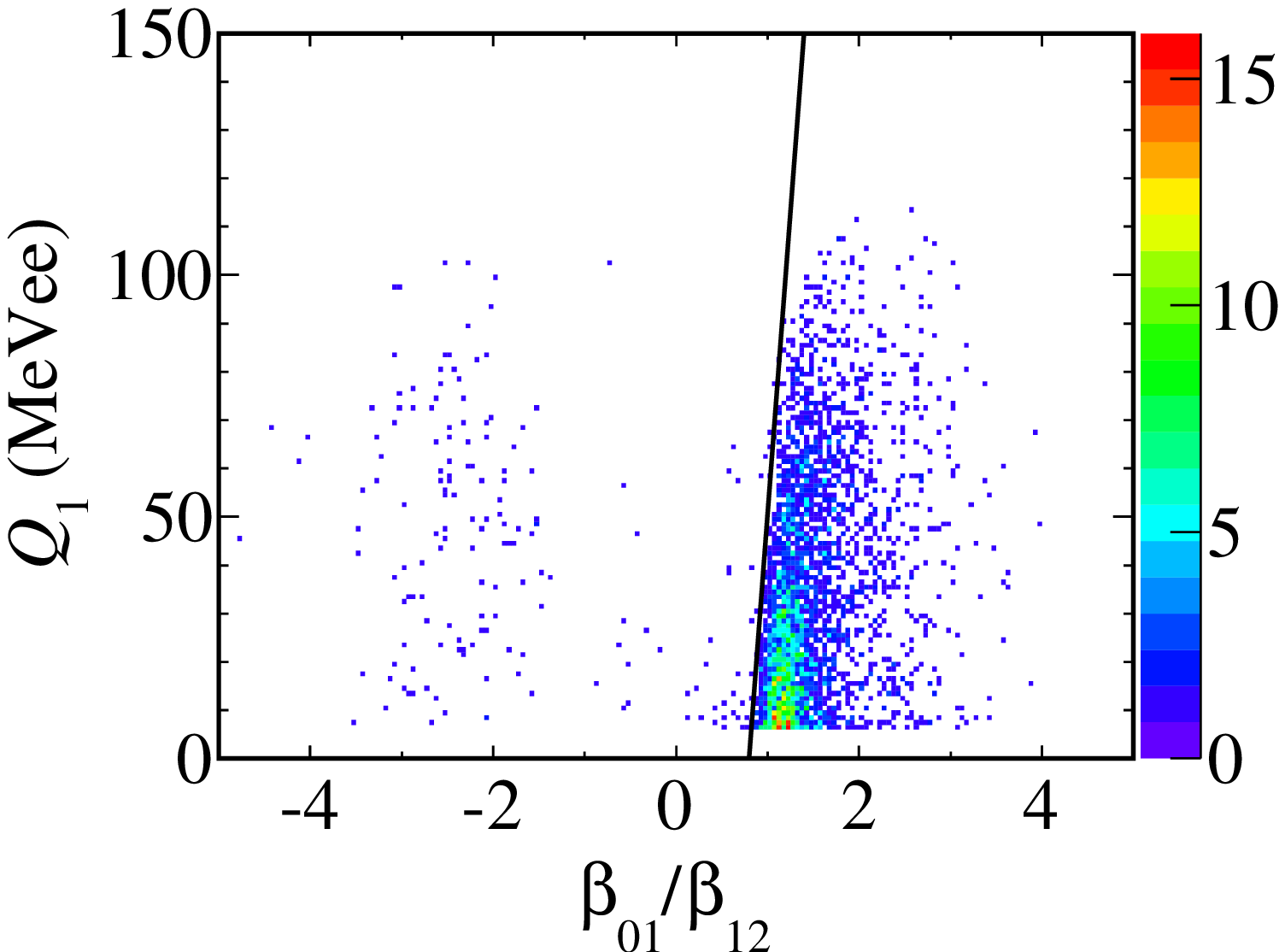,scale=0.38}
\epsfig{file=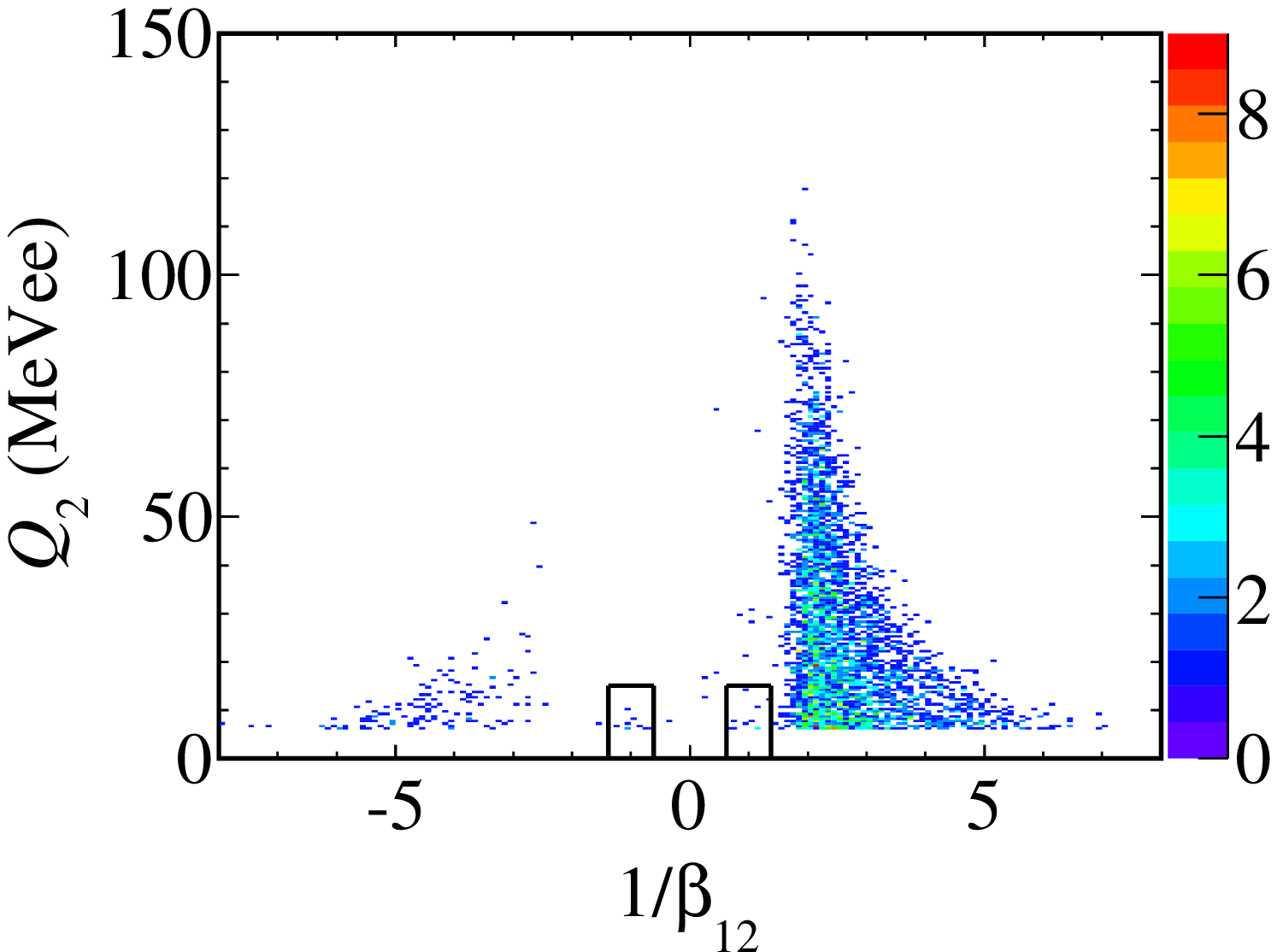,scale=0.38}
\end{minipage}
\begin{center}
\begin{minipage}[ht]{10. cm}
\caption{Spectra of the multiplicity $M\ge 2$ events observed in 
the $^7$Li$(p,n)^7$Be reaction. 
Left: $Q_1$(pulse height in the 1st wall) versus $\beta_{01}/\beta_{12}$.  
The right hand side of the line is caused by the event shown 
in Fig.~\ref{fig:cross_df_nebula}(a), while the events 
$-4 < \beta_{01}/\beta_{12} <-1$ correspond to Fig.~\ref{fig:cross_df_nebula}(b).
Right: $Q_2$(pulse height in the 2nd wall) versus $1/\beta_{12}$.  The squares represent
 for the cut for the $\gamma$-ray cross talk.
\label{fig:cross_df_nebula_data1}}
\end{minipage}
\end{center}
\end{figure}

The cross talks relevant to the different-wall 
events are schematically illustrated in Fig.~\ref{fig:cross_df_nebula}. 
The spectra of the cross talk events (multiplicity $M\ge2$ in the NEBULA modules) 
induced by single quasi-monoenergetic neutrons in the $^7$Li($p,n$)$^7$Be reaction 
at 200 MeV is shown in Fig.~\ref{fig:cross_df_nebula_data1}.
The energy threshold of the detectors was set 
to be 6 MeVee (electron equivalent) to remove most of the $\gamma$ rays produced
in the scintillator.

\begin{figure}[htb]
\begin{center}
\begin{minipage}[ht]{7. cm}
\epsfig{file=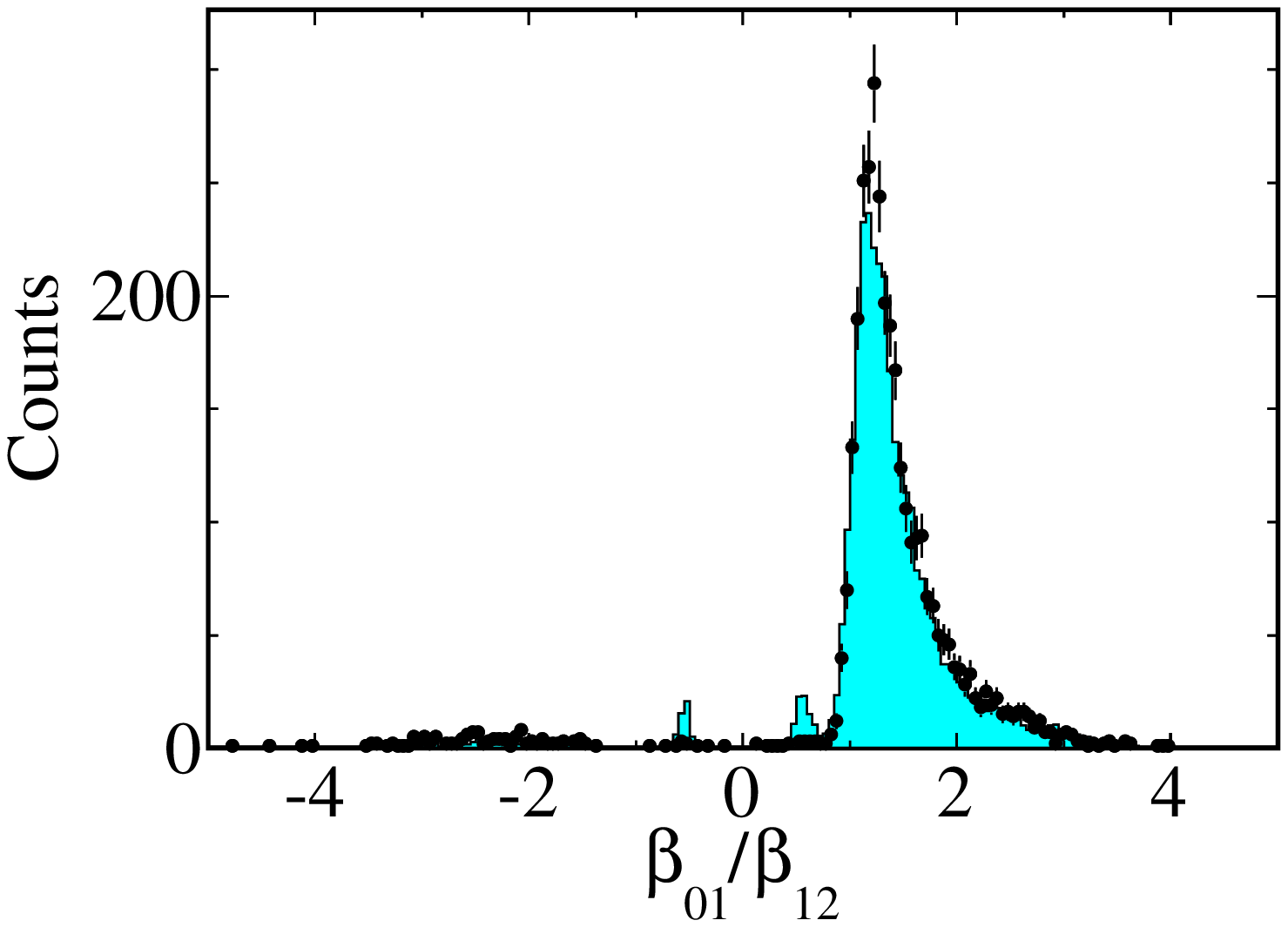,scale=0.45}
\end{minipage}
\hspace*{0.5cm}
\begin{minipage}[ht]{4. cm}
\caption{$\beta_{01}/\beta_{12}$ distribution (solid points) 
obtained in the $^7$Li$(p,n)^7$Be(g.s.+0.43 MeV) reaction, which 
is compared with the GEANT4 simulation (histogram). The small peaks at
$\pm\sim$~0.6 are due to $\gamma$ rays ($\beta_{12}=1, \beta_{01}=0.57$). 
\label{fig:cross_df_ratio}}
\end{minipage}
\end{center}
\end{figure}

The left spectrum is shown as a function of the charge 
$Q_1$ obtained in a module in the 1st wall
versus the velocity ratio $\beta_{01}/\beta_{12}$ (left), while the right figure
is shown as a function of the pulse height $Q_2$ obtained in a module in the 2nd wall versus 
$1/\beta_{12}$(right). Here $\beta_{01}$ represents the velocity between the target and
the first registered module, while $\beta_{12}$ is that 
between the first and the second registered modules.
If the first signal is registered in the 2nd wall, 
$\beta_{12}$ is negative.

The main cross-talk component, which lies to the right of the line
in Fig.~\ref{fig:cross_df_nebula_data1}~(left),
is due mainly to the quasi-free scattering on $^{12}$C, and
to the scattering of the neutron by a proton (hydrogen) in the scintillator material.
This corresponds to the event shown in Fig.~\ref{fig:cross_df_nebula}(a).
In this case, the neutron after the 1st wall is slower than the
neutron before the 1st wall, and thus $\beta_{01}/\beta_{12}> 1$.  The line in Fig.~\ref{fig:cross_df_nebula_data1}(left)
represents the boundary of this component, and is tilted 
since the more the energy that is lost in the first wall, 
the smaller the $\beta_{12}$ (and the larger the $\beta_{01}/\beta_{12}$). 
In the rejection procedure in an experiment involving two neutrons, 
the events on the right hand side of this line are eliminated.

On the other hand, there are much fewer events with $\beta_{01}/\beta_{12}< -1$,
which are interpreted as neutrons arising from neutrons evaporated from 
the 2nd wall.
This component corresponds to the event shown in Fig.~\ref{fig:cross_df_nebula}(b).
Such neutrons are expected, as observed in Ref.~\cite{IWAM11}, from the
interaction of high energy nucleons with the C nuclei. 

Figure~\ref{fig:cross_df_nebula_data1}(right) 
shows the spectrum of $Q_1$ vs. $1/\beta_{12}$,
which shows more clearly the events 
arising from $\gamma$ rays that traverse the two walls.
The squares shown in the figure are 
the conditions used to eliminate the $\gamma$-ray cross talk.
It should be noted that these $\gamma$-ray cross talk events
are produced in the detector material, 
and not caused by the reaction at the target. 

The projection onto the velocity ratio in the $^7$Li$(p,n)$$^7$Be reaction
is shown in Fig.~\ref{fig:cross_df_ratio}, 
compared with the GEANT4 simulation. As shown, the cross-talk events are
well reproduced by the simulation.

\subsection{Cross talk in same-wall events}

\begin{figure}[htb]
\begin{center}
\begin{minipage}[ht]{6. cm}
\epsfig{file=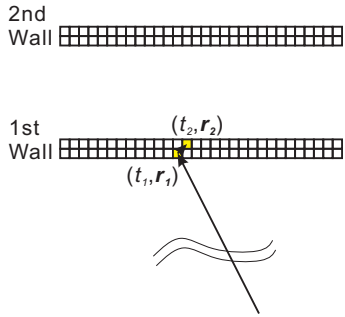,scale=0.5}
\end{minipage}
\begin{minipage}[ht]{4. cm}
\caption{Examples of cross-talk events relevant to same wall events.
\label{fig:cross_sm_nebula}}
\end{minipage}
\end{center}
\end{figure}

The cross talk relevant to
same wall events are schematically shown in Fig.~\ref{fig:cross_sm_nebula}.
In this case, the cross talk occurs mostly 
between neighboring modules. 

\begin{figure}[htb]
%\epsfysize=9.0cm
%\begin{center}
\begin{minipage}[ht]{11.5 cm}
\epsfig{file=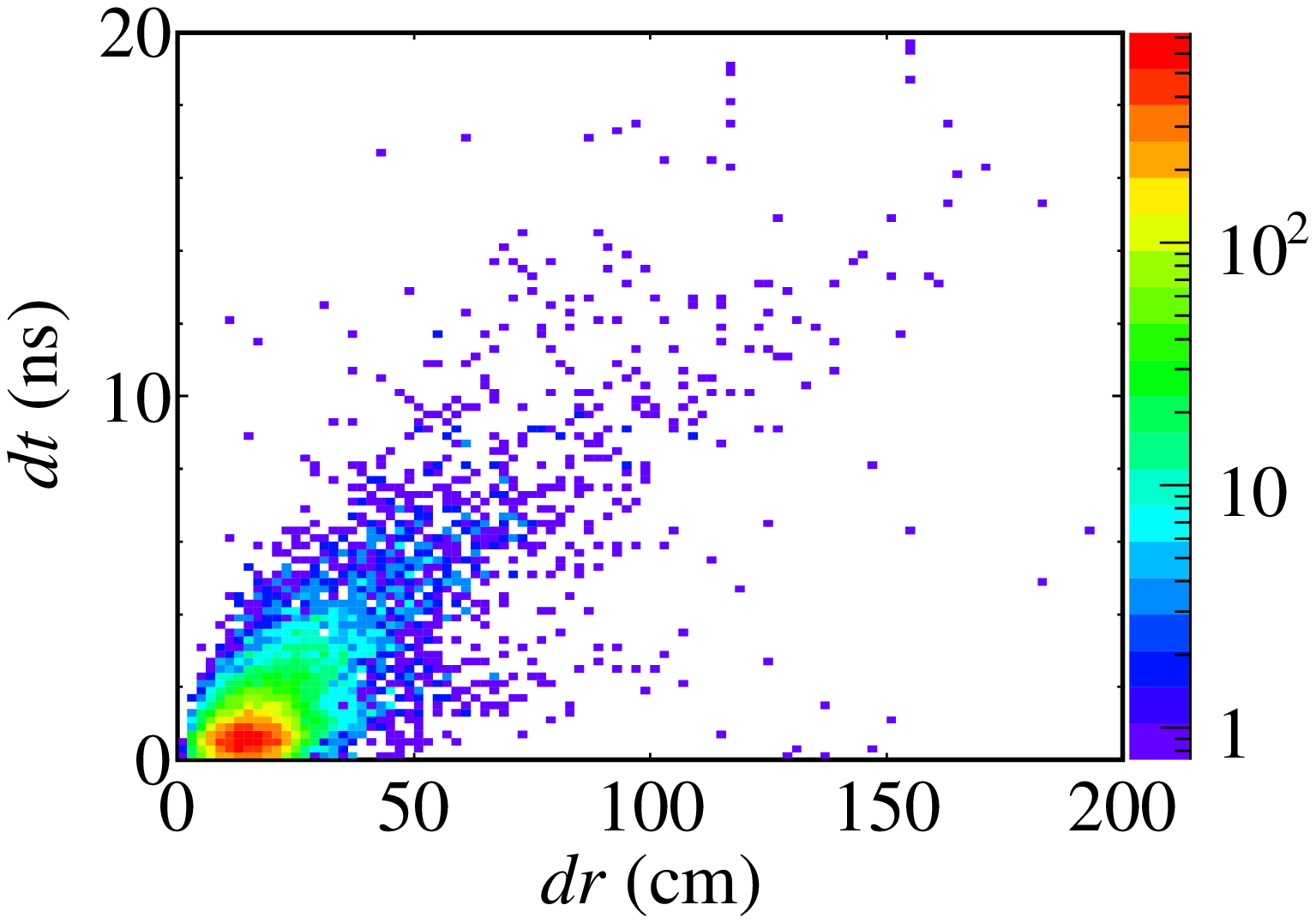,scale=0.38}
\epsfig{file=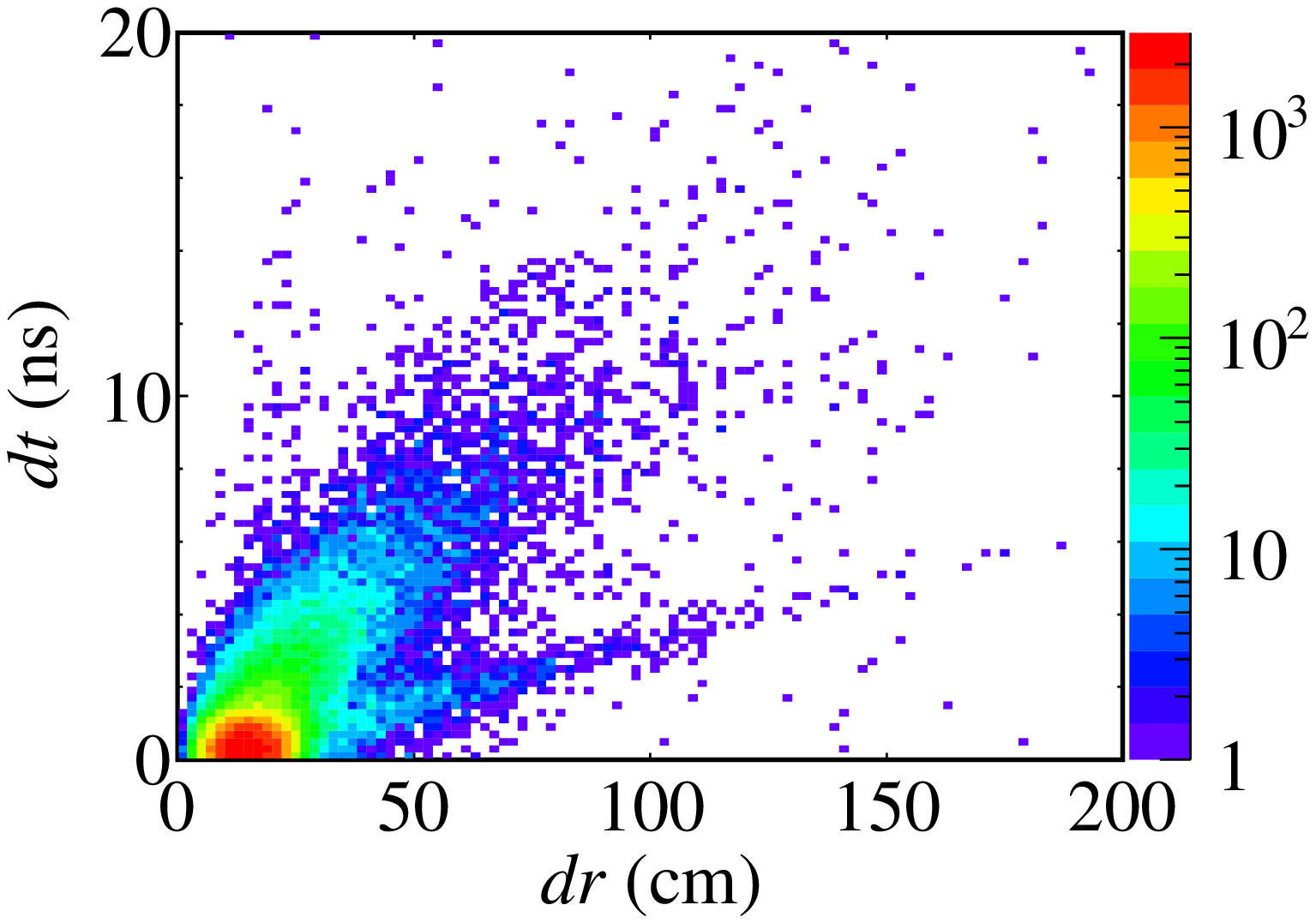,scale=0.38}
\end{minipage}
\begin{center}
\begin{minipage}[ht]{10. cm}
\caption{Plots of the two hit events (both in the 1st wall, or both in the 2nd wall) observed in 
the $^7$Li$(p,n)^7$Be reaction.
Left: Experimental spectrum of $dr$ versus $dt$.
Right: Results of the simulation. 
\label{fig:cross_same}}
\end{minipage}
\end{center}
\end{figure}

\begin{figure}[htb]
%\epsfysize=9.0cm
%\begin{center}
\begin{minipage}[ht]{11.5 cm}
%\begin{center}
\epsfig{file=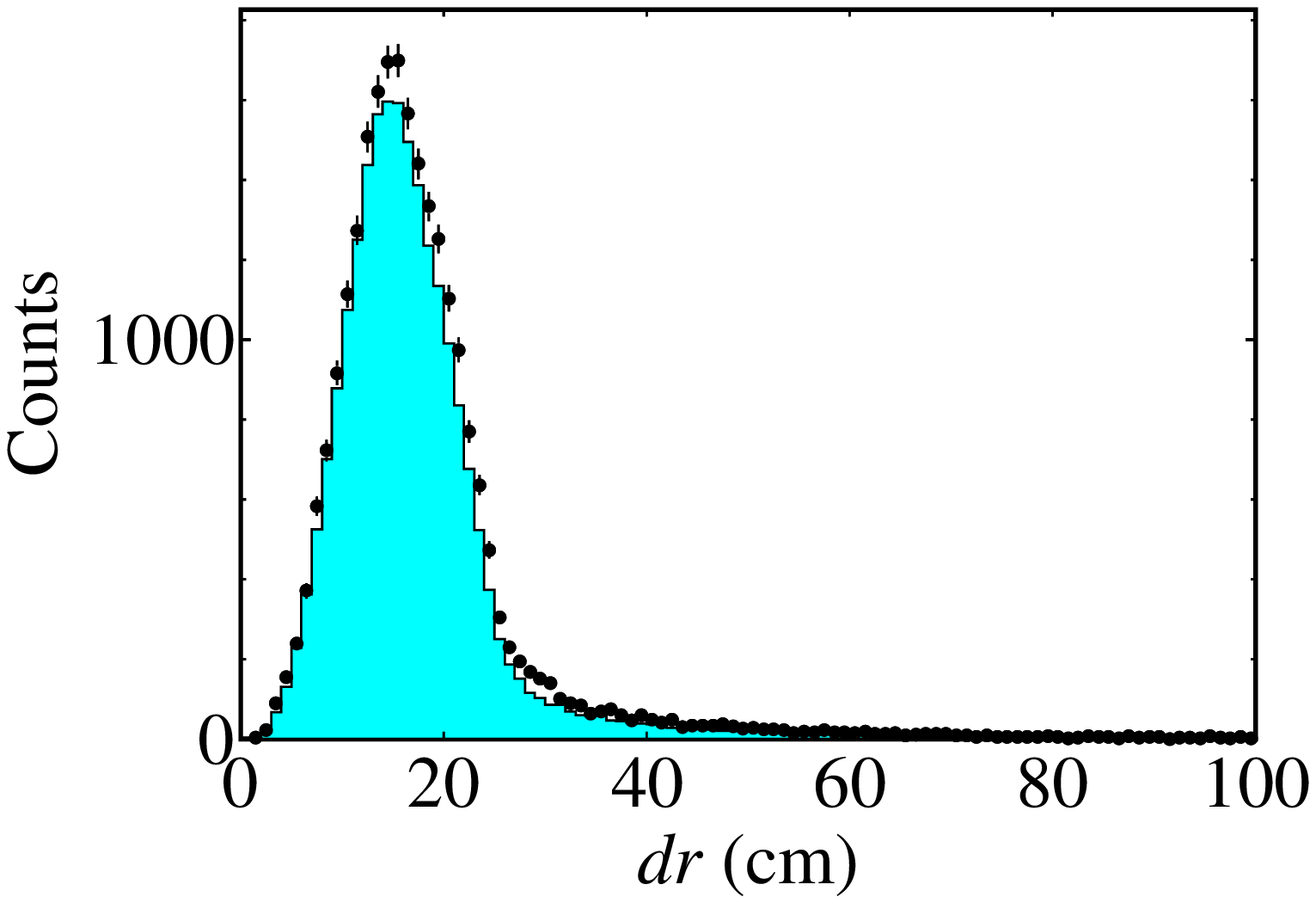,scale=0.38}
\epsfig{file=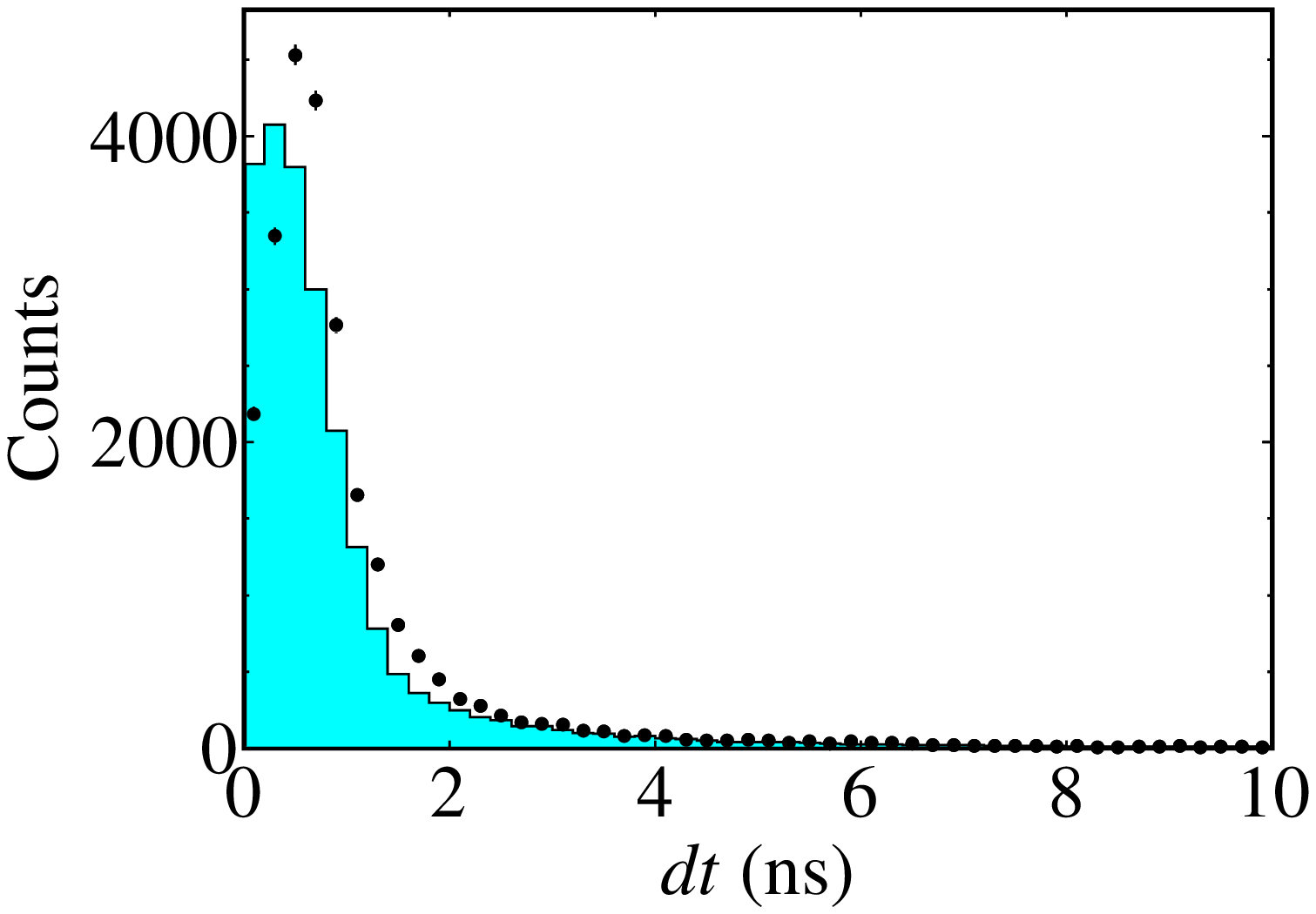,scale=0.38}
%\end{center}
\end{minipage}
\begin{center}
\begin{minipage}[ht]{10. cm}
\caption{Solid dots are experimental data of the distance (left) and the time difference 
between the two hits in
the $^7$Li$(p,n)^7$Be(g.s.+0.43 MeV) reaction, relevant to the cross-talks in the same wall.
Results from the simulation are shown by the solid histograms.
\label{fig:cross_same_comp}}
\end{minipage}
\end{center}
\end{figure}

Figure~\ref{fig:cross_same} shows spectra of 
the time difference ($dt$) against the distance ($dr$)
between the two hits in the same wall in the $^7$Li$(p,n)^7$Be data (left). 
Here, $dt=t_2-t_1$ and $dr=|\vec{r_2}-\vec{r_1}|$,
where $(t_1,\vec{r_1})$, and $(t_2,\vec{r_2})$ are respectively
the timing and three-dimensional coordinate of the two signals in the same wall 
caused by a single neutron. As shown the results of the simulation
are almost identical to the experimental data. 
The agreement between the data and the simulation is even more clearly seen in the projected $dr$ and $dt$
distributions shown in Fig.~\ref{fig:cross_same_comp}, demonstrating the validity of 
the simulation.
%As shown, the simulation reproduces well the experimental
%cross-talk events.

The cross-talk events 
are rejected using the condition,
\begin{equation}
\sqrt{ 
\left( \frac{dr-dr_0}{R} \right)^2 + \left( \frac{dt-dt_0}{T} \right)^2} <1,
\label{eq:same}
\end{equation}

where $dr_0=$15.8~cm, $R=15.0$~cm, $dt_0=0.50$~ns, and $T=18.3$~ns were determined empirically.
%When this condition is met, we judge this two-neutron hit as a cross talk (one-neutron hits
%two modules in the same wall). 
In this case, the first hit is adopted as a representative hit.
%In addition, we also use the $Q_2$ vs. $\beta_{01}/\beta_{12}$ to determine the further cut
%condition. With such conditions, the true two-neutron events in an experiment to measure two neutrons
%are also reduced at very low relative energies, since there are in the condition of Eq.~\ref{eq:same}.

\subsection{Evaluation of cross-talk cuts}

\begin{figure}[htb]
%\epsfysize=9.0cm
%\begin{center}
\begin{minipage}[ht]{11.5 cm}
\epsfig{file=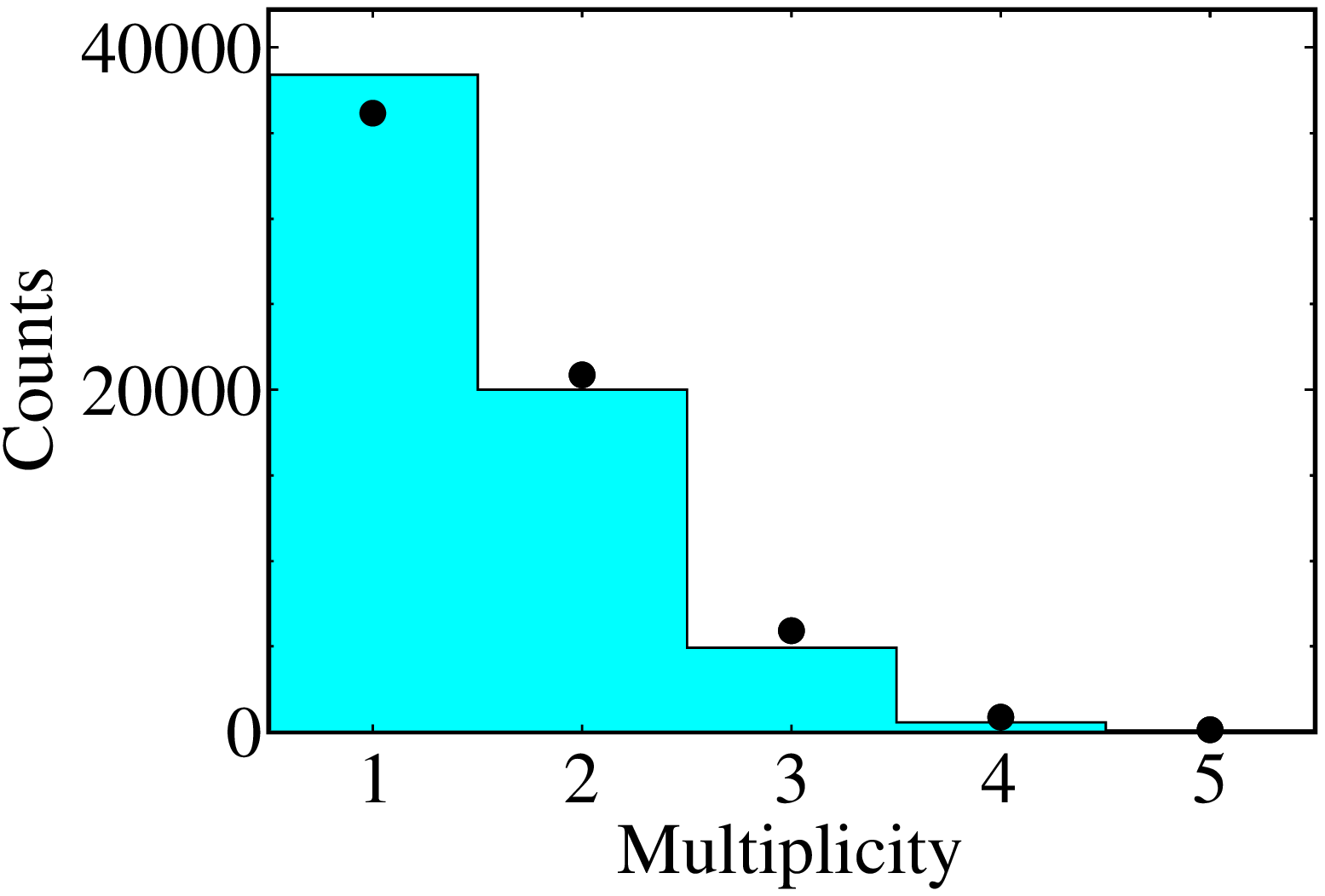,scale=0.38}
\epsfig{file=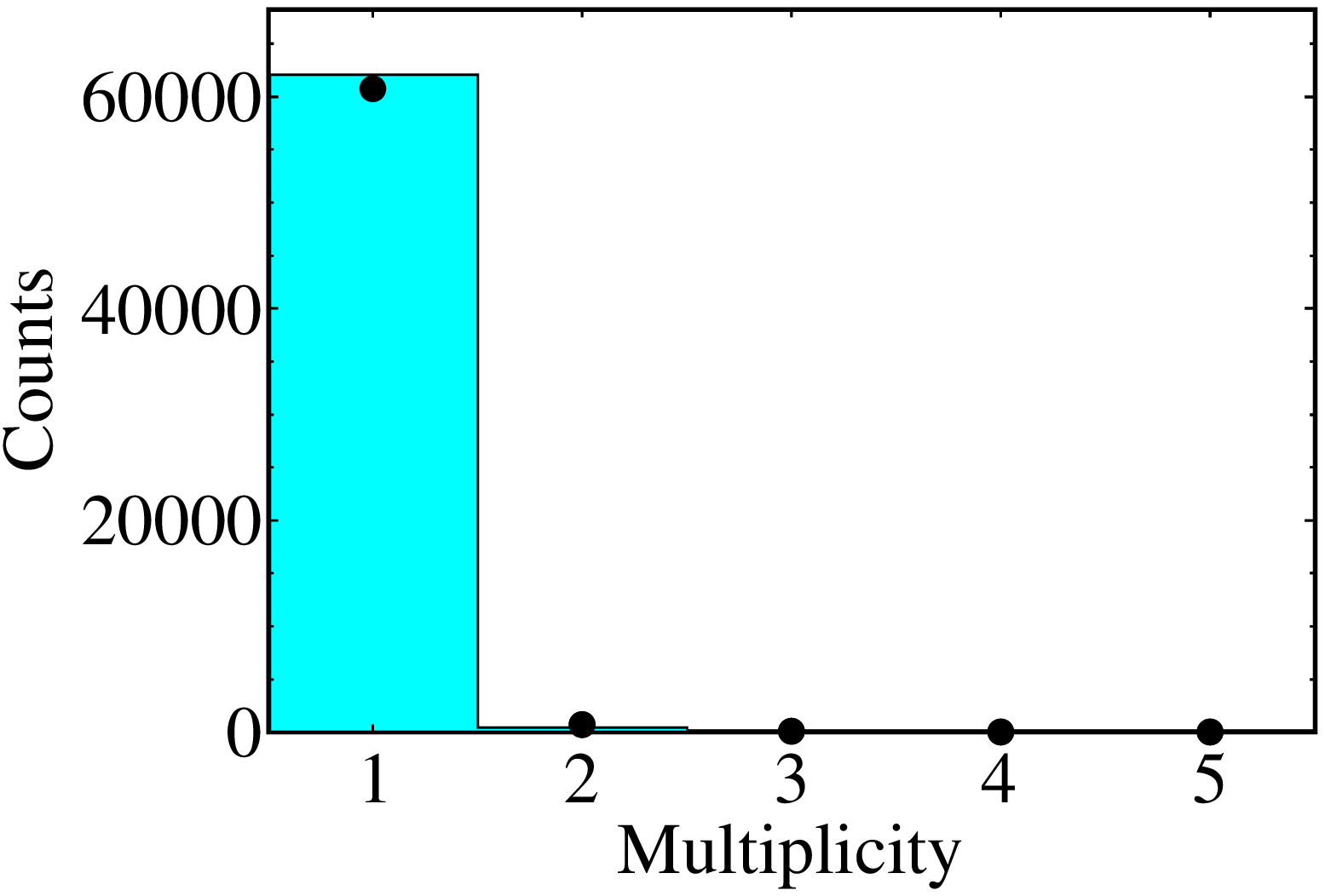,scale=0.38}
\end{minipage}
\begin{center}
\begin{minipage}[ht]{10. cm}
\caption{Multiplicity distribution obtained for the
$^7$Li$(p,n)^7$Be reaction before the cross-talk rejection (left) and
after (right). The solid dots are experimental results, while the
histograms are obtained from the simulation.
\label{fig:cross_mult}}
\end{minipage}
\end{center}
\end{figure}

Figure~\ref{fig:cross_mult} shows the multiplicity distribution
of the single neutron from the $^7$Li$(p,n)^7$Be reaction and the simulation,
before (a) and after (b) the cross talk elimination procedure.
Firstly, as shown, the experimental results are well reproduced by the simulation
both before and after the cross talk elimination.
Secondly, the vast majority of the cross talk is eliminated: most of the events
with $M\ge 2$ are either eliminated or summed up to the $M=1$ events.
More specifically, 97.1\% of the cross talk is eliminated
in the $^7$Li$(p,n)^7$Be data, while
98.4\% is eliminated in the simulation. 
%Nearly same evaluations for the elimination ratio for the data and the simulation
This further
demonstrates that the cross talk is well understood and the rejection
procedures are valid.

\begin{figure}[htb]
\begin{center}
\begin{minipage}[ht]{6. cm}
%\begin{left}
\epsfig{file=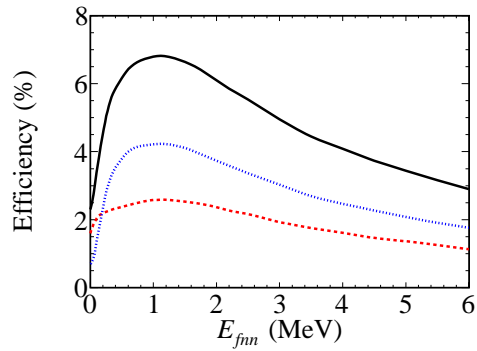,scale=0.42}
%\end{left}
\end{minipage}
\hspace*{0.5cm}
\begin{minipage}[ht]{4. cm}
\caption{Two-neutron detection efficiency obtained by the simulation for the case of
$^{26}$O$\rightarrow^{24}$O$+n+n$ produced by $1p$ removal from $^{27}$F at 201 MeV/u.
The different-wall events are shown by 
the red dashed line, while the same-wall events are the blue dotted line.
The sum is shown by the black solid line.
\label{fig:eff}}
\end{minipage}
\end{center}
\end{figure}

Finally, the efficiency for the two neutron detections are estimated as shown in Fig.~\ref{fig:eff}.
The very low $\erel$ less than 200 keV are unsurprisingly
more efficiently detected in different walls as the neutrons are emitted
with a small opening angle.
%since event the two neutrons in the same direction can be distinguished by the causality cut,
%which is very useful to probe the barely unbound states, such as in $^{26}$O~\cite{LUND12,CAES13,KOND15}
%or soft dipole excitation as in $^{11}$Li~\cite{NAKA06}.
On the contrary, the $2n$ efficiency drops rapidly for
the same wall events below 200~keV due to the cross-talk rejection procedure (Eq.~\ref{eq:same}).

%Meanwhile,
%the different-wall events are kept rather constant, which is suitable for probing the
%states very close to the decay threshold as in the case of $^{26}$O~\cite{KOND15}.

\section{Summary and Future prospects}

We have shown that large-acceptance spectrometers are
 a very useful tool for probing nuclei at the limits
of stability. Some key elements are: 
1) large acceptance for neutron detection, 
and 2) relatively high resolution
for clear mass identification. 
These features have been realized, as shown here, 
in the advanced large-acceptance spectrometer SAMURAI.
We have also discussed the issue of cross-talk when we need to measure multiple neutrons. 
% revised
In particular, we have shown that a multi-wall neutron detector array is well suited to such studies and provides good
detection efficiencies even at low $\erel$.

In the near future, at SAMURAI, 
it is planned to measure the 4$n$ decay of $^{28}$O.
To realize such an experiment,
the MINOS target~\cite{OBER11} and the NeuLAND neutron detectors have been added to the setup.
MINOS is a thick cryogenic LH$_2$(liquid hydrogen) 
target coupled to a light-particle tracker for the vertex reconstruction.
Hydrogen has the most atoms per gram, which is significant to 
maximize the yield of $^{28}$O. 
%for scarce four-neutron coincidence events.
Since MINOS can determine the vertex, 
the ambiguity of the energy loss in the target is reduced. Hence, relatively
good energy resolution is expected even with such a thick target ($\sim$ 15~cm). 
%Hence, the MINOS has the great potential, when it
%is combined with the large acceptance spectrometer as in SAMURAI.

Neutron detectors with large volume and high granularity 
is also important to measure more than two neutrons.
Due to the cross talk rejection procedures, 
the larger the number of separated walls the better the neutron detection efficiency.
With such a motivation, 400 NeuLAND modules ($5\times5\times250$~cm$^3$ each) 
have been installed at SAMURAI, in addition to the existing NEBULA detectors, for the next 2-3 years.
An upgrade of the NEBULA array, NEBULA-Plus
proposed by the LPC group (Orr {\it et al.}), has been approved, which will also facilitate
multi-neutron measurements.

In FRIB at MSU, the
HRS ({\bf H}igh {\bf R}igiditiy {\bf S}pectrometer) project 
has been proposed~\cite{BAUM15,ZEGE14}. 
As discussed here, one of the important requirements
of an advanced large acceptance spectrometer is a high momentum resolution 
for charged fragments, in particular to separate heavier masses.
The HRS will be equipped with focusing elements (quadrupoles)
which can provide for the high momentum resolution. 
A sweeper magnet with a large gap will also be developed,
to provide for a large acceptance for neutrons ($\pm 6^\circ$), 
when MoNA and LISA are used in the forward direction.

For the future FAIR facility, the development of R$^3$B (Reactions with Relativistic Radioactive Beams)
is underway. This includes 
the large superconducting dipole magnet 
(field integral $BL\sim 5$~Tm), and 3000
NeuLAND modules to realize a $4n$ detection efficiency of nearly 60\%~\cite{R3B}.
As mentioned, some of the first NeuLAND modules have been introduced
to RIBF, in advance of the experiments at FAIR.
Early physics runs with the R$^3$B setup at GSI are also expected in 2018.

Advanced large acceptance spectrometers are being built and developed world-wide.
Many more opportunities for physics studies are expected, as new associated
devices are added. As such, 
we expect that the large acceptance spectrometers will continue to play significant
roles in the next decade in RI-beam physics.

\section*{Acknowledgment}

The authors would like to thank the SAMURAI collaboration.
In particular, H.~Otsu, T. Kobayashi, N.A. Orr, J.Gibelin, M. Marques,
Y. Satou, M. Sasano, J. Yasuda, T. Isobe, T. Murakami, T. Motobayashi,
V. Panin, 
Y. Togano, S. Koyama, R. Tanaka, 
J. Tsubota, M. Shikata, T. Ozaki, A. Saito, 
K. Yoneda, H. Sato, T. Kubo, and T. Uesaka. 
We are grateful to M. Thoennessen 
and T. Aumann for giving T.N. slides and materials.

%\section*{References}

\bibliography{ref.bib}

\end{document}